\documentstyle[12pt,aaspp4]{article}

\begin{document}

\title{The Multiplicity of the Hyades \nl and its Implications for \nl Binary Star
Formation and Evolution}

\author{J. Patience\altaffilmark{1}, A. M. Ghez\altaffilmark{1,2,3}, I. N. Reid\altaffilmark{4}, \nl A. J. Weinberger\altaffilmark{4}, \& K. Matthews\altaffilmark{4}}

\vspace{5.25in}

\altaffiltext{1}{UCLA Division of Astronomy and Astrophysics, Los Angeles, CA
90095-1562}
\altaffiltext{2}{Sloan Fellow}
\altaffiltext{3}{Packard Fellow}
\altaffiltext{4}{Palomar Observatory, California Institute of Technology, Pasadena, CA 91125}

To appear in the Astronomical Journal, May 1998

\newpage
\begin{abstract}
A 2.2 $\mu$m speckle imaging survey of 167 bright (K $< 8.5$ mag) Hyades
members
reveals a total of 33 binaries with separations spanning 0\farcs044 to 
1\farcs34 and magnitude differences as large as 5.5 mag. Of these binaries,
13 are new detections and an additional 17 are now spatially resolved
spectroscopic binaries, providing a sample from which dynamical masses and 
distances can be obtained. The closest 3 systems, marginally
resolved at Palomar, were re-observed with the 10m Keck telescope in order to 
determine accurate binary star parameters. Combining the results of this 
survey with previous
radial velocity, optical speckle, and direct imaging Hyades surveys, the 
detected
multiplicity of the sample is: 98 singles, 59 binaries, and 10 triples.

A statistical analysis of this sample investigates a variety of multiple
star formation and evolution theories. Over the binary separation range 
0\farcs1 to 1\farcs07 (5 to 50 AU),
the sensitivity to companion stars is relatively uniform, with $<\Delta K_{lim}>$
= 4 mag, equivalent to a mass ratio $<q_{min}>$ = 0.23. Accounting for the 
inability to detect high flux ratio binaries
results in an implied companion star fraction ($csf$)
of 0.30 $\pm$ 0.06 in this separation range. The Hyades $csf$ is intermediate
between the values derived from observations of T Tauri
stars ($csf_{TTauri}=$0.40$\pm$0.08) and solar neighborhood G-dwarfs 
($csf_{SN}=$0.14$\pm$0.03). This
result allows for an evolution of the $csf$ from an initially high value 
for the pre-main sequence to that found for main sequence stars.

Within the Hyades, the $csf$ and the mass ratio distribution 
provide observational tests of binary formation mechanisms. The $csf$ is 
independent of the radial distance from the 
cluster center and the primary star mass.  
The distribution of mass ratios is best fit by a power law $q^{-1.3\pm0.3}$
and shows no dependence on the primary mass, binary separation, or the 
radial distance from the cluster center. Overall, the Hyades
data are consistent with scale-free fragmentation, but inconsistent with
capture in small clusters
and disk-assisted capture in small clusters. Without testable
predictions, scale-dependent fragmentation and disk fragmentation
cannot be assessed with the Hyades data.

\end{abstract}

\newpage
\section{Introduction}

Early surveys of solar neighborhood stars found that binaries outnumber
solitary stars (Abt \& Levy 1976), and more recent results have reinforced the
observation that multiples are very common (Duquennoy \& Mayor 1991). 
Surveys of T Tauri stars, young objects only a few
million years old, revealed a surprisingly large fraction of binary stars,
with twice as many companions compared to solar neighborhood G-dwarfs
over a semi-major axis range from $\sim 10$ to 250 AU (Ghez, Neugebauer \&
Matthews 1993; Leinert et al. 1993; Simon et al. 1995; Ghez et al. 1997).
This discrepancy suggests the possibility of a decline in the overall
binary frequency with time.
Since the Hyades has an age ($\sim 6 \times 10^{8}$yr) between the T Tauri 
stars and the solar neighborhood,  and also has a nearby distance 
(D = 46.3 pc, Perryman et al. 1997) and carefully determined membership, 
the cluster is well-suited for an investigation
of the evolution of the companion star fraction. 

Young clusters are also ideal laboratories for binary star formation
studies since they provide a sample of stars with 
relatively constant age, metalicity, and distance. Binary star formation 
models fall into
two broad categories: capture and fragmentation (cf. Clarke 1996). 
Capture has been postulated to proceed in small-N (4-10 member) clusters either
with or without the dissipative effects of disk interactions (McDonald \&
Clarke 1993, 1995). The restriction to small-N clusters rather than large 
clusters 
is necessary because the probability of an interaction that forms a binary is 
too low in large clusters (Clarke). Alternatively, different models of
fragmentation have also been proposed -- fragmentation of the protostellar 
cloud core or of the circumstellar disk (cf. Boss \& Myhill 1995; Myhill \& 
Kaula 1992; Burkert \& Bodenheimer 1996; Bonnell \& Bate 1994). Observable 
properties 
of binary stars, such as the distribution of mass ratios and the dependence 
of the companion star fraction on the primary star mass, provide important
tests of binary star formation scenarios.

In this paper, the results of a speckle imaging survey of the Hyades are 
presented. The separation range covered, $0\farcs10$ to
$1\farcs07$ (5 to 50 AU), not only fills the gap between spectroscopy
and direct imaging, but also overlaps the $\sim30$ AU peak of the 
distribution of semi-major axes measured for 
binary stars (Duquennoy \& Mayor 1991; Mathieu 1994). The main goal of the project
is to conduct a statistical analysis of the properties of the
observed binary stars in order to test the predictions made by binary
star formation and evolution scenarios.
The membership and magnitude criteria used to 
select the sample are explained in \S 2. Section 3 describes the
 observations, followed by the details of the data analysis procedures 
presented in \S 4. The results of the survey and the bounds of the 
completeness region are given in \S 5, which also includes a comparison 
of the present survey with the considerable amount of previous work on the 
Hyades. In the 
discussion, \S 6, the observed binary star properties are analyzed in order
to explore theories of
binary star formation and  evolution. 
Finally, the main conclusions are summarized in \S 7.

\section{Hyades Sample}

The stars selected for this speckle survey satisfy both a magnitude limit
of $K<8.5$ mag and a membership 
requirement based on proper motion and photometry. The 
number of stars satisfying the membership and magnitude criteria is 197, and 
167 of these stars were observed; the observed sample is listed in Table 1.
The four red giant stars and the evolved A star in the Hyades were observed 
and the results are reported, but the majority of the statistical analysis
is confined to the main-sequence stars. 
This speckle sample represents approximately one-third of the total
cluster census. The target list of Hyades members was culled from the 
appendix of Reid (1993) 
which identifies probable members on the basis of proper motions and optical 
photometry. Because only the brighter members of the cluster are included in 
the
 speckle survey sample, most of the astrometric and photometric observations 
are from either the original investigation of the structure and motion of the 
Hyades by van Bueren (1952) or from the subsequent Leiden photographic survey 
by  Pels et al. (1975). 

Candidates identified in these early studies have 
been subjected to more recent and more selective membership tests with 
highly accurate proper 
motions (Schwann 1991), photometry (Mermilliod 1976),
and astrometry (Perryman et al. 1997). The Hipparcos satellite measured the
parallax to 139 Hyades stars in the speckle survey. Combining the 
results of the parallax measurements of many Hyades members provides the 
most accurate distance to the cluster center, 46.3 pc (Perryman).
The individual proper motions, however, have a smaller uncertainty
($\sim$2, \% Schwann) than the individual Hipparcos parallaxes 
($\sim$10, \% Perryman et al.). The smaller uncertainties make the proper motion data
more sensitive to the relative distance between members. Because of these
considerations, the distance to each star listed in Table 1 is determined by
scaling the value given in Schwann (1991) by 0.966, the ratio of the distance 
to the cluster center measured by Hipparcos and proper motions.
The 
distances for 25 of the faintest stars not measured by Schwann were scaled from
the appendix in Reid (1993) which lists the result from the Pels et al. study.

Although optical photometry has been obtained from previous membership studies,
K-band photometry is not available for most of these stars. 
An estimate of the K magnitude of each star
was obtained by combining the V magnitude and B-V color listed in Reid (1993) 
with an empirical color-color transform described in the Appendix.
At the 46.3 pc mean distance 
to the Hyades, the limiting magnitude corresponds to a minimum target
star mass of 0.46 $M_{\odot}$, based on the mass-M$_K$ relation also 
given in the Appendix. The depth of the cluster, 15\%, causes 
a variation of at most 0.05 $M_{\odot}$ in the target mass limit.

\section{Observations}

Speckle observations of the Hyades stars were obtained at a wavelength
of 2.2 $\mu m$ between 1993 and 1996 at the Cassegrain
focus of the Hale 5 m telescope with the facility near-infrared camera. 
Over the three-year period that the 
observations were made, this instrument was upgraded; the camera array was 
replaced once and the reimaging optics which  determine the pixel scale were 
changed twice. Table 2 summarizes the details of each observing run. Each night
approximately twenty stars were observed, and 
during the last two nights, 26 stars with an initial $\Delta K_{lim} < 3.0$
(cf. \S 4) were reobserved to improve 
the data quality. Additionally, the 3 marginally-resolved binaries 
(vB 91, vB 96, and +10 568) were 
reobserved on 1996 December 22 and 1997 December 14 with the W. M. Keck 10 m 
telescope and its speckle imaging system (Matthews et al. 1996).

For each target star, a total of 3,000 to 4,000 exposures of $\sim$0.1
seconds 
were recorded. These source observations were interleaved with similar 
observations of a reference 
point source in sets of $\sim$500 images. The short exposure time is 
necessary to  ``freeze'' the turbulent 
structure of the atmosphere, and a large number of 
images provides many samples of the instantaneous effects of the
atmosphere, as required for speckle imaging.
The rapid exposure permitted 
the use of the broad band K-filter ($\Delta \lambda = 0.4 \mu m$) for most of the 
observations,
however, six of the brightest stars -- vB 28, vB 41, vB 70, vB 71, vB 8, and
vB 33 -- were observed through a one-percent cvf 
filter centered on a wavelength of 2.2 $\mu m$ in order to prevent the
array 
from saturating. Since the central wavelength of the cvf filter is identical 
to that of 
the K-band, the resolution of these six observations is comparable to that
of 
the other stars in the sample. During one night, poor seeing conditions
allowed
similarly bright sources to be observed through the K-filter. 

\section{Data Analysis}

The initial data reduction steps 
follow standard image analysis -- the raw speckle images are sky subtracted,
flat fielded, and corrected for dead pixels by interpolating over neighboring
pixels. The subsequent steps follow the 
method developed by Labeyrie (1970) to compute the square of the Fourier 
amplitudes
$|\widetilde O(f)|^2$ for each star. Binary stars are differentiated from 
single stars by their distinct pattern in the power spectrum $|\widetilde O(f)|^2$; single stars
exhibit a uniform $|\widetilde O(f)|^2$, 
while a binary system displays a periodic $|\widetilde O(f)|^2$ given by

\begin{equation}
|\widetilde O(f)|^2 ={R^2 + 1 +2Rcos[2\pi \vec {\theta} \cdot \vec f] \over R^2 + 1 + 2R} ,
\end{equation}

\noindent
where R is the flux ratio and $\vec {\theta}$ is the 2-dimensional separation
on the sky. 
For stars with the characteristic sinusoidal fringe pattern of a 
binary, a chi-squared minimization of a two-dimensional model fit to the 
Fourier amplitudes provides estimates of the flux ratio, separation, and 
position angle ($\pm 180^\circ$) of the binary (Ghez et al. 1995).  
The remaining $180^\circ$ ambiguity in the position angle is 
eliminated by determining the Fourier phases as prescribed by Lohmann et al.
(1983). 
Examples of two speckle binaries with different
separations and position angles are shown in Figure 1.

{\centering \leavevmode
\epsfxsize=.45\columnwidth \epsfbox{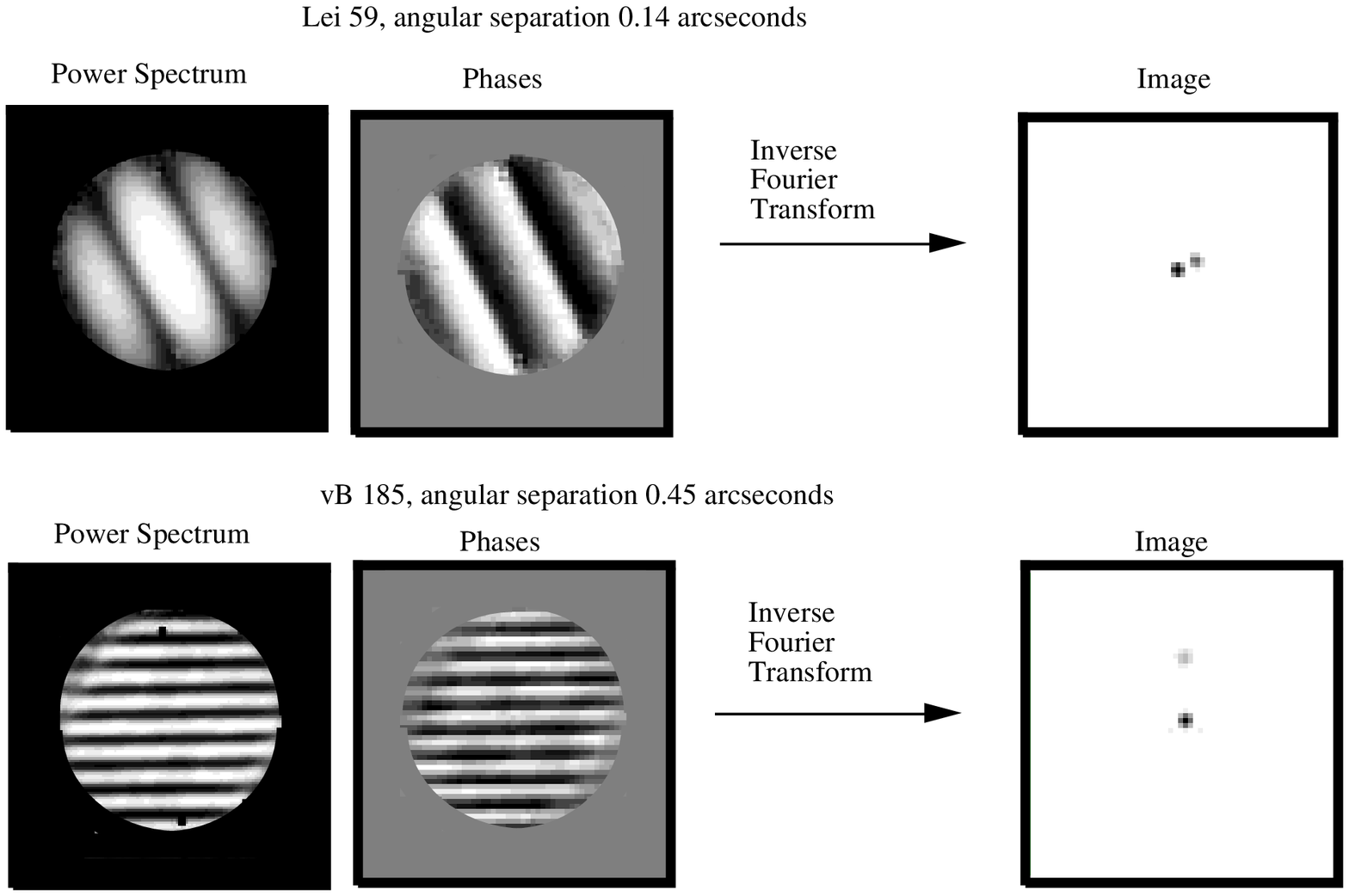}
\hspace*{.08\columnwidth}
\parbox[b]{.45\columnwidth}{Figure~1: Two example speckle imaging reconstructions are shown for the 
Hyades binaries vB 24 and 
vB 185. The left fringe pattern displays the calibrated Fourier
amplitudes $|\widetilde O(f)|^2$, and the right fringe pattern displays the 
Fourier phases $(arg\widetilde O(f))$.
Using an inverse Fourier transform, the diffraction-limited image of
each binary is produced.}}

The separation and flux ratio are well-determined by the model fit 
for systems in which the first minimum 
occurs at a spatial frequency less than D/$\lambda$, equivalent to a 
separation greater than $\lambda$/2D ($i.e. > 0\farcs05$). Three stars -- 
vB 91 vB 96, and +10 568 -- which have formal separation solutions of 
exactly the theoretical resolution limit in the Palomar data, 0\farcs05, 
may actually have slightly 
different separations, since the data does not show if the decline in the 
Fourier amplitudes extends entirely to the first minimum. These stars were 
reobserved with the 
10m Keck telescope in 1996 and 1997, one to three years after the initial Palomar 
observations and the more accurate flux ratios determined
from the Keck data are used to refine the Palomar measurement of the 
separations. The separation of vB 91 and vB 96 stars increased in the 1996 data, 
probably as a 
consequence of orbital motion. The Keck measurements are reported in Table 3, 
but the statistical analysis described in $\S 6$ is restricted to the Palomar 
results in order to maintain a consistent set of data.

For the widest binaries, the magnitude difference estimate from the 
model fit method is overestimated; this occurs because 
some of the flux in the speckle
 cloud of wide companions falls outside the array and because the images are 
apodized before their power spectra are computed. To avoid this 
bias, the binaries with separations wider than one-half the field of 
view minus 
one-half the speckle cloud size ($i.e. >0\farcs70$) are reanalyzed with the shift 
and add technique (cf. e.g. Bates \& Cady 1980; Christou 1991).
The systems analyzed with this technique are vB 40, vB 17, vB 151, Lei 52,
Lei 130, +22 669, vB 52, and vB 5, as noted in Table 3. As expected, 
the shift and add 
$\Delta K$ values are all smaller than the results from the speckle analysis, 
although the differences between the two methods are only significant for 
separations larger than 1$\farcs$0. 

The final step in the data analysis computes the limits for possible unseen 
companions to the single stars. These limits vary with atmospheric conditions,
the target star brightness,
and the distance from the target star.
The companion detection limits, $\Delta K_{lim}$, of each single star 
observation is found by solving for the 
maximum amplitudes of several cosine waves corresponding to a separations of 
0\farcs05, 0\farcs06, 0\farcs07, 0\farcs10, 0\farcs15, and 0\farcs60  
that could be hidden in 
the noise of the Fourier amplitudes. The method is similar to that described in 
Ghez et al. (1993) and Henry (1991), but the maximum simulated cosine waves are 
only allowed to vary from 3 times the rms scatter to unity rather than from the
lowest power spectrum value to unity.

\section{Results}
\subsection{IR Speckle Results}

Of the 167 stars observed, 33 are resolved as 
binary systems; nearly half of the binaries, 12 systems, are new detections. 
The properties of the speckle binaries are listed in Table 3 and each binary is
plotted in Figure 2. The smallest separation measured was $0\farcs044$, and the 
largest K magnitude 
difference measured was 5.5 mag, which corresponds to
a companion mass of only 0.10 M$_\odot$ and a mass ratio of 0.13 
(see Appendix). The faintest companion has an apparent K magnitude of 12.8.
Each detected pair is assumed to be bound, since the probability of a chance 
superposition is only $\sim {.01 \%}$ 
given the $\sim {4 \times 10^{-5}}$ per square arcsecond
surface density of field stars with
K $< 12$ in the direction of the Hyades (Simon et al. 1992). 

{\centering \leavevmode
\epsfxsize=.45\columnwidth \epsfbox{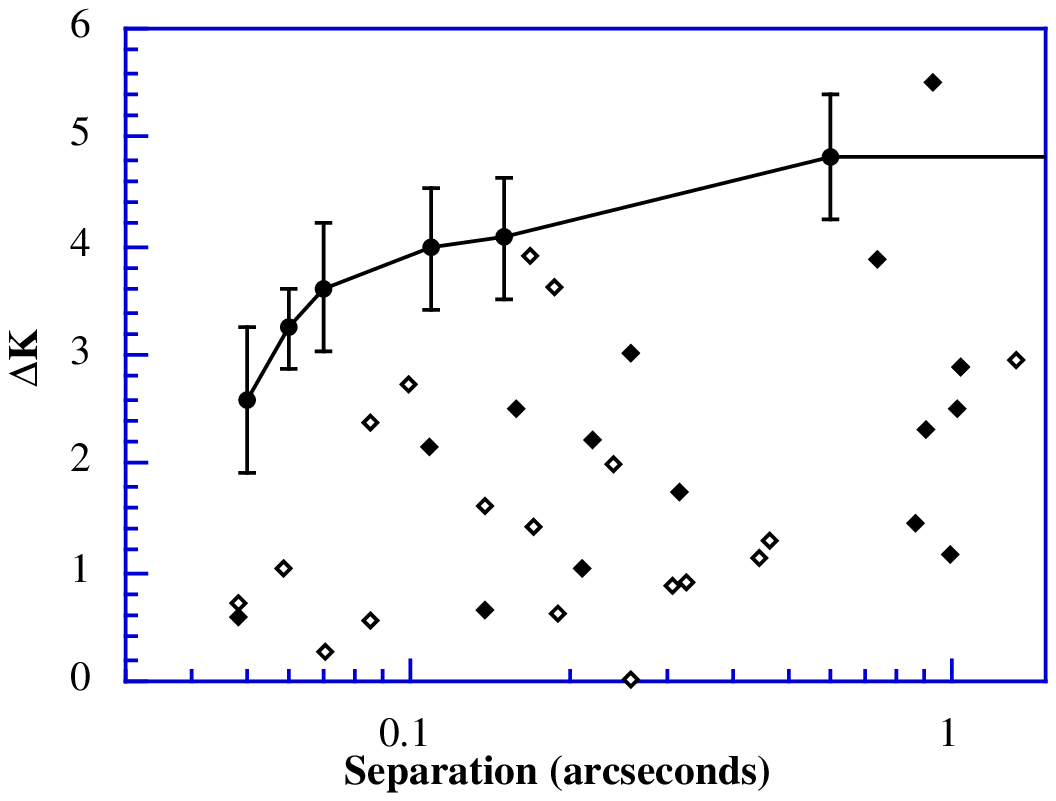}
\hspace*{.08\columnwidth}
\parbox[b]{.45\columnwidth}{Figure~2: The observed properties for the 33 detected binaries (new
binaries -- filled diamonds, known binaries -- open 
diamonds) are 
compared to the median companion star detection limits (filled 
circles) at several separations. The error bars are the standard 
deviations of the limits. At the completeness limit of the survey,
0\farcs10, the observations are, on average, sufficiently sensitive to detect a
binary star with a magnitude difference of $\Delta K=4$.}}

Before discussing the statistical properties of the Hyades binaries in
the speckle sample, an accurate accounting of the 
sensitivity of the observations is needed to define the separation and 
$\Delta K$ parameter space over which the survey is complete -- the 
``completeness region.'' The upper limit on the projected
separation range is set by the camera field-of-view. With the target star 
image centered on the array, the upper 
cutoff is $1\farcs07$, one-half of the field-of-view with 
the finest pixel scale. The smallest angular separation reliably measurable 
with two-dimensional speckle imaging, with the assumption that
the object Fourier amplitudes follow a binary star cosine pattern, 
 is $\lambda$/2D, or $0\farcs05$. Although three binaries are resolved  
very close to this limit, many observations lack the sensitivity to 
detect companions at this extreme (see Table 4). The average
sensitivity limit as a function of separation is shown in 
Figure 2, where the error bars represent the 1-$\sigma$ rms variations in 
the sensitivity limits; these values are based on the $\Delta K_{lim}$ 
computed for the single stars.
The completeness region lower cutoff is chosen to be $0\farcs10$ in order 
to maintain a nearly uniform sensitivity to companions at all separations. 
At this lower cutoff for the separation 
range the median $\Delta K_{lim}$ is  4.0 mag. At the distance of the Hyades, 
the angular separation range of the completeness region corresponds to a 
projected linear separation 5 to 50 AU. A total of 7 of the 33 pairs 
are detected outside the separation range of the completeness region and are
therefore not included in the complete sample. Six of the 
binaries -- vB 57, Lei 90, vB 91, +10 568, vB 120, and vB 96 -- are omitted
because their projected separations are less than the lower limit cutoff,
while vB 40 is excluded from the complete sample because it has
a separation larger than the upper limit cutoff. 

In summary, over the binary star projected separation range of 0\farcs10 
to 1\farcs07, the median of the detection limits is $\Delta K_{lim} =$
4.0 mag. At the distance of the Hyades, the angular separation 
range corresponds to a projected linear separation 5 to 50 AU. Based on the 
empirical mass-M$_K$ relation described in the Appendix, the median
magnitude difference limit corresponds to a median mass ratio limit of 0.23.
The derived detection limits and
companion star masses are plotted in Figure 3 as a function of the target 
star mass. For the lowest mass stars in the sample, the median detection limit
corresponds to companions of $\sim 0.2M_{\odot}$ -- within 
$\sim$0.1 $M_{\odot}$ of the hydrogen-burning
limit. The higher mass stars, however, typically have detection limits 
that only extend to $\sim 0.6M_\odot$ -- comparable to
the primary mass of the fainter stars in the survey. 

{\centering \leavevmode
\epsfxsize=.45\columnwidth \epsfbox{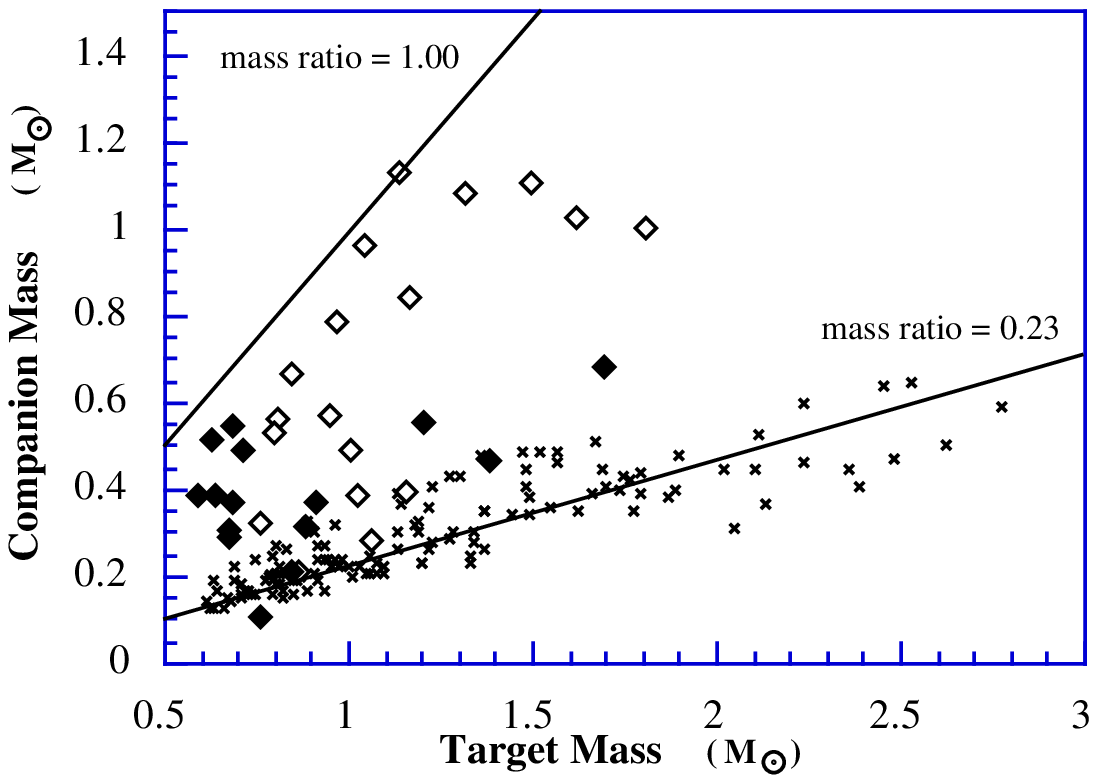}
\hspace*{.08\columnwidth}
\parbox[b]{.45\columnwidth}{Figure~3: The derived primary and secondary masses for the speckle binary stars (new
binaries - filled diamonds, known binaries - open diamonds) are plotted
along with the companion star mass detection limits for the stars
observed as single in the speckle survey
(x). Although mass ratios as large as 0.13 are observed, the median mass 
ratio cutoff for this survey is 0.23.}}

\subsection{Comparison with Previous Surveys}

Previous investigations of the cluster multiplicity have utilized optical 
speckle, spectroscopy, and direct imaging, and in this section the results of several such 
studies are compared to this project (see Table 3 \& 4 notes). The optical 
speckle survey by Mason et al. (1993) includes most of our
targets, with 133 stars in common. Twenty-eight of these 133 stars have been 
spatially resolved as binaries: 11 by both
surveys, 4 stars -- BD +35 714, vB 71, vB 41, and vB 132 -- 
by Mason et al., and 13 by the current infrared (IR) survey alone. 
The 60\% higher binary detection
rate at IR wavelengths results from the enhanced sensitivity to smaller
mass ratio main sequence binaries at longer wavelengths (see Appendix). The
optical speckle survey detection limit, $\Delta V_{lim} = 3$, corresponds to a mass ratio of only 0.54, a factor of $\sim 2$ less 
sensitive in mass ratio than the present IR speckle observations.
Of the 4 stars missed by this IR study, 2 are easily explained.
Both vB 71 and vB 41 are giant stars which, unlike main sequence stars,
require a larger dynamic range to observe 
a companion in the IR than at optical wavelengths. Additionally,
the separation for vB 71, $0\farcs048$, is slightly below the limit 
observable with the present IR survey.
The discrepancy with the stars +35 714 and vB 132 may result from either
significant orbital motion or a $\Delta K$
greater than the limit listed in Table 3. The latter alternative seems 
unlikely, since the
$\Delta K_{lim}$ of 4.5 and 3.1 in the current data implies $\Delta V$
detections greater than 7.0 and 4.7, both of which are beyond the
detection limit of the optical speckle results. 
Orbital motion, however, could position the companion star closer 
to the primary 
than the current IR speckle resolution in the three years between the 
optical and infrared measurements.

Repeated spectroscopic observations of many Hyades stars have been made by 
several authors (e.g. Griffin et al. 1988; Stefanik \& Latham 1992); in general, 
radial velocity measurements 
detect short-period binaries unresolvable with
speckle imaging. Nonetheless, many of the longer period 
spectroscopic binaries can, in principle, be spatially resolved.
A 3 yr orbit represents the shortest-period 
orbit resolvable with this speckle survey, assuming a total system mass of 
$\sim 1M_{\odot}$
and the extreme conditions of an eccentricity near unity and a face-on 
orbit observed to have an angular separation of 0\farcs1 at 
apastron.
The minimum detectable period increases to $\sim$9 yr
for a circular orbit. Because of the incomplete overlap in separation range 
covered by speckle and spectroscopy, stars observed as binaries by both 
techniques can
be either triples for which each technique detects a different
pair of stars in the multiple system, or doubles for which the same pair is
detected. The notes in Table 3 indicate which speckle binaries have also 
been measured spectroscopically and which binaries are actually triple systems
with separate speckle and spectroscopic pairs.

The common sample between 
the IR speckle survey and the Griffin et al. (1988) radial velocity survey 
contains 87 stars. Of the 33 Griffin et al. binaries in this set,
15 are also resolved by the speckle measurements. The separation
and period are so discrepant for 8 of the 15 speckle/spectroscopic binaries -- Lei 20, 22 669, vB 40,
Lei 83, vB 124, vB 185, vB 102, and vB 151 -- that they must be triple stars
consisting of a spectroscopic binary and a third star orbiting further away.
The remaining 7 binaries resolved by both surveys -- vB 57, 
vB 113,
Lei 90, vB 96, vB 114, vB 59, and vB 91 -- are systems for which the two
techniques are probably detecting the same pair. Five spectroscopic systems --
vB 115, Lei 57, vB 106, vB 71, and vB 39 -- have periods $>$3 yrs, but were not resolved by the IR 
speckle survey; vB 71 has already been discussed (see comparison with 
optical speckle), and the rest may be at orbital positions which 
correspond to a separations
below the resolution limit of the IR speckle measurements or they may be single-lined binaries with faint companions.

An additional 11 spectroscopic binaries
in the speckle sample are listed in Stefanik \& Latham (1992) and 6 of 
these systems were resolved
with the speckle observations. All 6 binaries resolved with speckle -- 
vB 24, vB 29,
vB 58, vB 75, vB 122, and vB 124 -- have orbital periods consistent with the 
observed speckle separations, so it is unlikely that any are triple stars.
Despite the long period of vB 131, it was not resolved. The four remaining
systems have such short periods that they are unresolvable with these 
speckle observations, and one of the short-period binaries -- vB 34 -- 
also has a white dwarf companion (Bohm-Vitense 1993), making it a triple 
system. A second star in the speckle 
sample has a white dwarf companion -- +16 516 (V471 Tau Nelson \& Young 1976).
Another 8 spectroscopic systems are listed in Barrado y Navascues \& 
Stauffer 
(1997). Although the orbital periods are not given, 4 of the spectroscopic
binaries -- vB 81, vB 50, vB 52, and vB 120 -- were resolved with speckle, 
and are assumed to be double, not triple, stars. 
Five additional early-type spectroscopic systems with known periods are noted 
in Table 4, and the periods are given in Abt (1965), Abt \& Levy (1985), and
Burkhart \& Coupry (1989).
The 17 pairs detected by both speckle and spectroscopy
provide a rare opportunity to accurately determine the 
mass and distance of each star without relying on additional 
assumptions about 
the stars or the Hyades cluster. Two such studies have already
been carried out for vB 57 and vB 24 (Torres et al. 1997a,b). 

Although more than one-half the current sample has been studied recently with 
spectroscopy, current direct imaging surveys of 
the Hyades have concentrated on the lower luminosity stars beyond the
magnitude limit of the speckle survey. For example, the imaging survey by 
Macintosh et al. (1997) includes only 39 of the stars in the 
speckle sample. Four of the stars in common -- vB 99, vB 105, vB 109, and 
vB 7 -- had candidate companions, but their large angular separations make
it unlikely that any of the pairs are physically associated. Another direct 
imaging survey of the Hyades involving 
HST observations does not include any of the stars in this survey (Gizis \& 
Reid 1995; Reid \& Gizis 1997).
Early photographic surveys
were capable of detecting bright (B $<12$) companions at modest
separations ($>5 \arcsec - 10$ \arcsec) among the brighter stars, but these
stars are heavily saturated on deeper plates. Thirty-six of the 
stars in the speckle survey are listed in the ADS or IDS catalogues as
visual doubles or triples with separations ranging from 0\farcs1 to 
88\farcs6 (Aitken 1932; Jeffers, van den Bos \& Greeby 1963). The eight 
systems in these catalogues with separations less than 1\farcs5 --
vB 29, vB 40, vB 57, vB 58, vB 75, vB 122, vB 124, and vB 132 -- have been
resolved with either optical or infrared speckle and are close enough to 
be considered physically associated (see \S 5.1). Most of the wider
``companions,''
however, are not Hyades members and are therefore discounted. Of the 7
visual binaries for which both 
stars are definite Hyades members -- vB 1/vB 2, vB 71/vB 72, vB 83/vB 182, 
vB11/vB 12, vB 54/vB 55, vB 56/vA 354, and vB 131/vB 132 -- only the 
$2\farcs0$ ($\sim$100 AU) vB 11/vB 12 
system is 
considered a binary in the analysis of binary statistics. It is unlikely that 
the other 6 systems are physically associated because either the distance to each 
star is different by more than 3.5 pc (3 $\sigma$) or the projected separation 
exceeds 4200 AU, the scale length between cluster members (Simon 1998).

Without considering the incompleteness of the different surveys, the
total number of binary or multiple systems detected by spectroscopy, speckle,
or direct imaging is 98 singles, 59 binaries, and 10 triples among the 167 
stars including the evolved stars. After 
considering the results from other techniques, the 33 speckle
binaries are actually 25 binaries and 8 triples. Similarly, the 134 speckle
singles become 98 singles and 34 binaries and 2 triples after including the 
other multiplicity data. Among the 
Hyades triple systems, all are hierarchical. The triple with the most similar 
separations is vB 102 with a 731 day period spectroscopic pair and a third 
star resolved by speckle at a distance of 0\farcs24, implying a ratio of 
semi-major axes of $\sim 8:1$ 

\subsection{Improved Color-Magnitude Diagram}
Unresolved binary stars significantly broaden the width of the 
main sequence, limiting the effectiveness of the color-magnitude diagram in
studies of age variations and rapid rotation.
With the combined data sets from radial velocity, speckle, and direct imaging,
the color-magnitude diagram of the Hyades cluster
can be improved by purging binaries from the graph. Since the widest binary
has a separation of only $2\farcs0$, all companions are close enough to
affect the photometry of the primary star. In addition to the effects of 
unresolved companions, the $\sim2\%$ uncertainty in the 
distance measurements also contributes to the spread within the color-magnitude 
relation.
Figure 4a shows the 167 stars in this sample
including the measurements of the known multiples, and Figure 4b
plots a noticeably narrower main sequence with only those stars with no 
known companions. Two stars remain significantly above 
the main sequence in Figure 4b -- vB 60 and +13 647 -- and are most 
likely unresolved binaries, although they are not counted as binaries
in the analysis that follows. Neither of these sources has a reported 
spectroscopic measurement.

Excluding the two giants and the two stars above the main sequence in
Figure 4b, the polynomial fit to the single star main sequence is 

\begin{equation}
M_V = -0.16 + 9.0(B-V) - 2.6(B-V)^2.
\end{equation}

\noindent
The standard deviation of the difference between the measured $M_V$ and the
$M_V$ expected from equation 2 is 0.12 for Figure 4a, half the scatter of 0.25 measured for Figure 4b. Although the Hyades cluster
is too old to place a meaningful limit on the age spread, a similar reduction
in the width of the main sequence will be important to constraining an age
spread in younger clusters. 

{\centering \leavevmode
\epsfxsize=.45\columnwidth \epsfbox{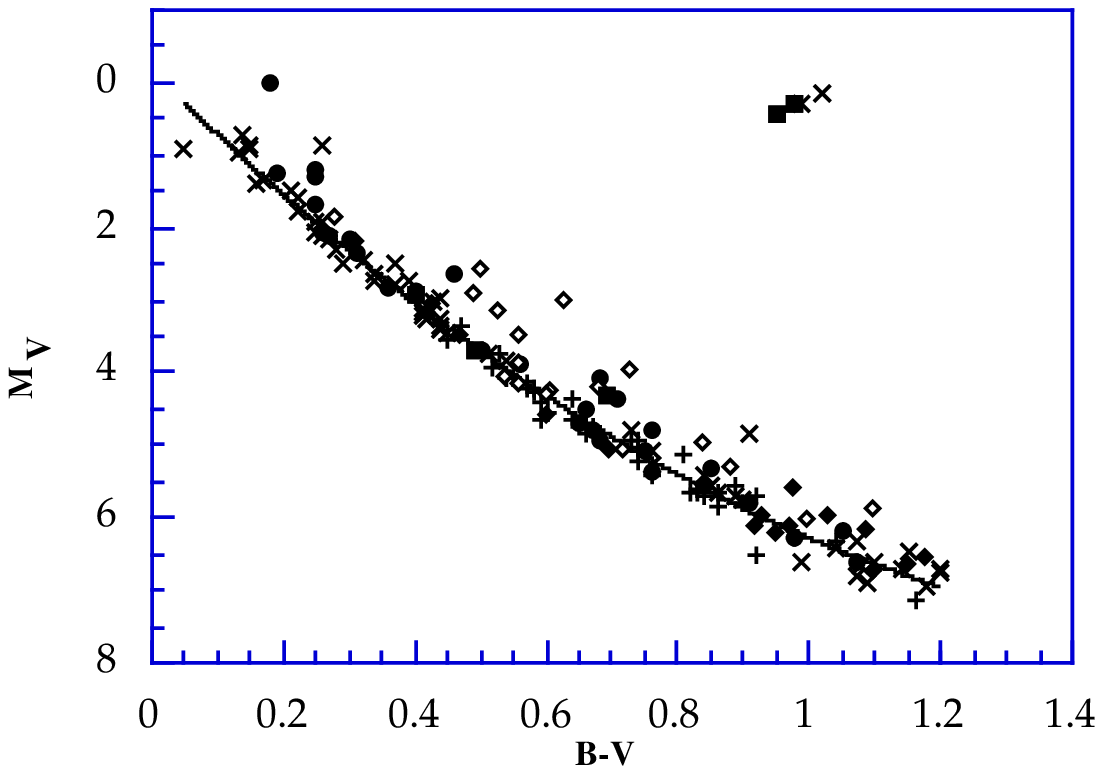}
\hspace*{.08\columnwidth}
\parbox[b]{.45\columnwidth}{Figure~4a/b: Color-magnitude diagrams are shown for (a) all of the stars in the 
sample and (b) only those stars with no known companions. The symbols for 
binary stars are: filled circles -- spectroscopic systems, filled squares --
optical speckle system or visual binary, and diamonds -- IR speckle binaries 
as in the previous figures. }}
{\centering \leavevmode
\epsfxsize=.45\columnwidth \epsfbox{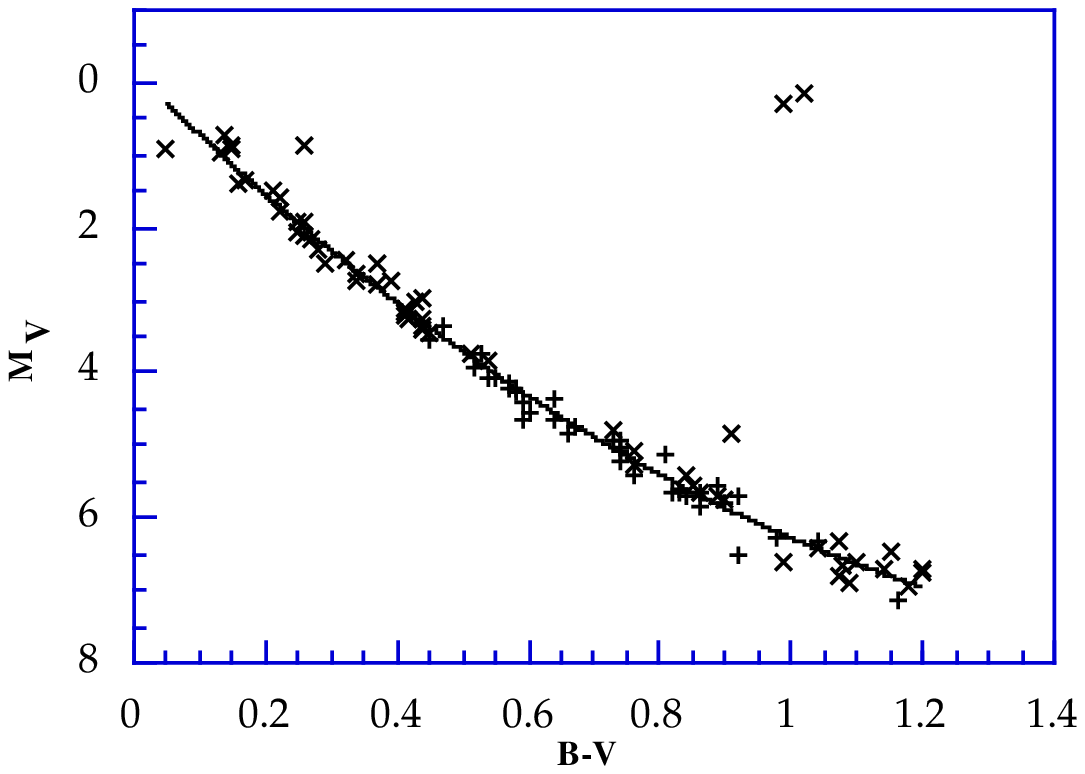}
\hspace*{.08\columnwidth}
\parbox[b]{.45\columnwidth}{Either the single stars have been 
observed with both speckle and spectroscopy (+), or they have been observed 
only
with speckle (x). The width of the main sequence is reduced by a factor of 2 
in Figure b, and the two stars located well above the main sequence in 
Figure b are probably photometric binaries.}}

\section{Discussion}
The following subsections examine the observed stellar properties in order to 
test the predictions of possible binary star formation and evolution scenarios. 
In $\S6.1$ the companion star fraction ($csf$) of the sample is calculated and its 
dependence on radius ($\S6.1.1$), mass ($\S6.1.2$), and time ($\S6.1.3$)
are compared to theoretical models. 
The calculations in the discussion involving the $csf$
consider all binaries detected with separations from $0\farcs10$ to 
$1\farcs07$. 
The mass ratio distribution and its radial and
mass dependence are calculated in $\S6.2$. The discussion describing the mass 
ratio distribution is also based on binaries in the $0\farcs10$ to 
$1\farcs07$ separation range, but the mass ratio range is restricted to 0.30 or larger.
With the $csf$ results, the mass ratio distributions are used to test several 
formation mechanisms. 

\subsection{Companion Star Fraction}
The current sample covers a well-defined range of separation
(5 to 50 AU) and mass ratio ($\sim$0.2 to 1.0), providing an excellent 
basis from
which the multiplicity of the Hyades can be determined. The number of 
companions can be quantified in two ways, the multiple star fraction or
the companion star fraction. The multiple star fraction,

\begin{equation}
msf = {b + t \over s+b+t},
\end{equation}

\noindent
does not differentiate between different order multiple systems while the
companion star fraction,

\begin{equation}
csf = {b + 2t \over s+b+t},
\end{equation}

\noindent
counts the total number of pairs, where s, b, and t are the number of 
singles, binaries, and triples. The uncertainties in the $csf$ and $msf$
are given by the Poisson counting error. 

A lower limit on the Hyades main sequence
multiplicity can be determined by combining the companions detected by the 
IR speckle survey with the
additional companions discussed in \S 5.2.
Outside the separation range of the speckle survey
it is difficult to gauge the completeness, and no corrections are applied
to account for undetected companions.  Spectroscopic, 
speckle, and direct imaging surveys have revealed 96 singles, 57 binaries, 
and 9 triples among the current sample, excluding the giant stars.
Given the total sample size of 162,
the $msf_{total,obs}$ is 0.41 $\pm$ 0.05, and the $csf_{total,obs}$ is 0.46 
$\pm$ 0.05. Because 
nearly one-half of the sample may not have been observed 
spectroscopically and all techniques have a limited sensitivity, the 
$msf_{total,obs}$ and $csf_{total,obs}$ are lower limits to the actual 
values. 

Within the restricted separation range of $0\farcs10$ to $1\farcs07$ (5 to 50 AU),
the observed $csf_{5-50AU,obs}$ and $msf_{5-50AU,obs}$ 
are  0.16 $\pm 0.03$ (26 of 162 
main sequence stars); the fractions are the same since no triple stars were
resolved.
Again, this value represents a lower limit on the total multiplicity between 
5 and 50 AU, since faint companions are not detectable.
The number of companions that lie within the separation range of this survey,
but at magnitudes below our detection limit, is estimated by assuming that the 
companion K luminosity distribution follows an observed K luminosity 
distribution. 
The hypothesis that the magnitude distribution of companion stars resembles 
that of single stars is supported by the solar neighborhood G-dwarf survey
which includes stars similar in mass to the Hyades survey (Duquennoy \&
Mayor 1991). 
The K-band luminosity function defined by all stars within 8 pc of the
Sun is used to model the Hyades companion star distribution fainter than
the detection limit (Reid \& Gizis 1997, Henry \& McCarthy 1992). The field 
luminosity function was selected instead of the Hyades K-band luminosity 
function because the observed population 
of Hyads is incomplete for the faintest stars, due to the greater difficulty 
in detecting these stars at further distances, and, possibly, due to the
evaporation of the lowest mass stars from the cluster (Reid 1993). 
Table 5 lists, for a given $M_K$, the 
percentage $p$ of 
the field sample with fainter magnitudes, 
and the number $N$ of speckle single stars in 
the Hyades sample with detection limits, $M_{Klim}$,
from $(>M_K - 1)$ up to and including $(M_K)$. The 
percentage of the main sequence that is undectable is determined by the 
average incompleteness,

\begin{equation}
 incompleteness = {\sum_{M_K=0.0}^{M_K=12.0} p N \over \sum_{M_K=0.0}^{M_K=12.0} N } .
\end{equation}

\noindent
For the 162 main sequence stars, the average percentage of the main sequence
that is undetectable is 46\%; dividing both the observed 
$csf_{5-50AU,obs}$ of 0.16 
$\pm$ 0.03 and the uncertainty by 0.54 to account for missing stellar 
companions yields a 
$csf_{5-50AU,corr}$ of 0.30 $\pm$ 0.06  over the projected 
separation range of 5 to 50 AU. Since the
total Hyades sample is divided into several subsamples in the following
sections, the detection limit groupings for 
each subsample are also listed in Table 5 as is the final assessment of the 
subsample incompleteness. In the following discussion, these subsamples 
are used to study binary star formation mechanisms and possible
evolutionary effects.

\subsubsection{Radial Distribution of Multiple Systems -- Imprint of Star 
Formation or Ongoing Relaxation?}
Evidence of mass segregation, a concentration of the higher mass stars at the
cluster center, has already been observed in the Hyades (Reid 1992),
and a similar segregation of the binaries is expected if the former 
result is due to cluster relaxation. Alternatively, high mass stars may
preferentially form at the cluster center as a consequence of enhanced 
accretion occurring in the region of highest gravitational potential 
(Bonnell et al. 1997; Zinnecker 1982). If the second scenario is responsible 
for the mass
segregation, then the binaries are not expected to be concentrated toward
the central region of the cluster. 

To investigate the radial distribution of multiple stars in the Hyades,
the multiplicity inside and outside of 3 pc are compared. The dividing radius
is chosen to be 3 pc since the mass function for the main sequence sample 
inside this radius is significantly different from the mass function outside 
this radius -- evidence of mass segregation. The coordinates 
given in Gunn et al. (1988) are taken as the center of the Hyades cluster and
a distance of 46.3 pc to the center is assumed.
The results, listed in Table 6, show no 
difference between the central and outer binary fraction of either the complete
speckle sample or the total binary/multiple sample which incorporates several 
techniques; varying the 
dividing radius does not alter the result. Although this result is consistent with
the competitive accretion model, the statistical significance of the conclusion
is low given that the secondary stars add an average of only 40 \% to the 
total mass of the system. A larger sample 
size would improve the significance of this conclusion. The lack of a radial 
dependence in the $csf$ is, however, consistent with the observed mass segregation
in clusters that are sufficiently young that dynamical evolution cannot have 
caused the higher mass stars to migrate toward the center (e.g. Hillenbrand 1997,
Sagar et al. 1988).

\subsubsection{Mass Dependence of the Companion Star Fraction -- an 
Observational Test of Scale-Free Fragmentation and Small-N Capture}
Certain binary star formation models predict distinct mass dependences for
the companion star fraction; scale-free fragmentation models produce binaries
with properties that are independent of the primary mass (Clarke 1998), 
while capture in small 
clusters preferentially forms binaries among the highest mass stars (McDonald
\& Clarke 1995). Based on theoretical calculations by Clarke designed for 
comparison to data sets with a constant mass ratio cutoff, the speckle
sample $csf_{obs}$ should
be independent of primary mass in the case of scale-free fragmentation, whereas
the same $csf_{obs}$ should increase with increasing primary mass for 
capture in 
small-N (4-10 star) clusters. To test the predictions of the two models, the 
sample is
split in two by B-V color in increments of 0.10 and the $csf$ for the 
stars bluer 
and redder than the cutoff is determined. For all B-V 
cutoffs, the bluer (higher mass) stars
have a consistently smaller $csf$ than the redder (lower mass) stars, although
the difference is never statistically significant. 
Table 6 lists the $csf_{obs}$ and $csf_{corr}$ for three B-V ranges
(used in \S 6.1.3) for both
the complete speckle binary sample and the total binary/multiple sample. 
The $csf$ of the more massive stars is not larger than that of the less 
massive stars, contradicting the expectation of the small-N capture model.
Although the paucity of substellar
companions detected in large surveys (e.g. Nakajima et al. 1995, Macintosh et al. 1997,
Zuckerman \& Becklin 1992) suggests that binary 
formation is not entirely scale-free, the results from this survey support 
the scale-free fragmentation model of formation for stars in the mass range
of the survey, $\sim 0.6 M_{\odot}$ to $2.8 M_{\odot}$.

\subsubsection{Evolution of the Companion Star Fraction}
The companion star fraction has been observed to differ significantly between
the pre-main sequence and main sequence stage of stellar evolution, with a 
larger proportion of binaries among the younger population. One proposed 
explanation for this discrepancy is the disruption of 
primordial multiple star systems over time which could be reflected in an
intermediate $csf$ for the Hyades sample (Ghez et al. 1993).
Among the alternate explanations are: an
environmental effect involving the different types of star forming regions and
a result of the shape of the evolutionary tracks which map a wider range of 
companion masses into a given detection limit at the pre-main sequence stage
(Ghez 1996). Ideally, any comparison
between samples of different ages is made over a common range of 
separation and sensitivity.
For this study, 5 to 50 AU defines the separation range, and mass ratios
from $\sim 0.2$ to 1.0 set the limits of the sensitivity range. 

A comparison set of the pre-main sequence binaries is taken 
from both lunar occultation and speckle surveys of T Tauri stars in the 
Taurus and Ophiuchus
star-forming regions (Ghez et al. 1993; Leinert et al. 1993; 
Simon et al. 1995). Since the nearest 
star-forming regions are three times as distant as the Hyades, the 
combination of lunar occultation and speckle ensures
that the entire 5 to 50 AU separation range is covered. 
Because K magnitudes do
not uniquely determine the mass of a T Tauri star, the 
$csf$ of the T Tauri stars is calculated by grouping the 
observations by their detection limits and then dividing the number of 
binaries with a certain range of flux ratios by the number of observations 
with the sensitivity to detect a companion
in that flux ratio range (cf. Ghez et al. 1997). The resulting companion star
fraction for the $\sim 2$ Myr-old pre-main sequence sample is 
$csf_{5-50 AU,T Tauri}$ = 0.40 $\pm$ 0.08. 

The older, $\sim 5$ Gyr-old comparison sample is taken from the multiplicity 
survey of
the solar neighborhood G-dwarfs (Duquennoy \& Mayor 1991). These data cover
10 orders of magnitude in orbital period, but the range 3.7 to 5.2 
logP[dys] corresponds to a projected linear separation range of 5 to 50 AU,
assuming a system mass of 1.4 $M_{\odot}$ (the average value for the G-dwarf 
sample) and a factor of 1.26 between the projected separation and 
the semi-major axis (Fischer \& Marcy 1992). Although this period range 
encompasses the 
results of two observing techniques, spectroscopy and direct imaging,
used in the G-dwarf survey, the majority
of this range is covered by direct imaging. The G-dwarf visual binary companion
correction limit of $\Delta V = 7$ mag
is comparable to the median Hyades limit of $\Delta K = 4$ mag 
(see Appendix). The $csf$ for 
the older solar neighborhood (SN) sample was calculated by
integrating the Gaussian fit to the corrected numbers of pairs
in the G-dwarf survey over the period range 3.7 to 5.2 logP[dys], 
yielding a $csf_{5-50 AU,SN}$ of 0.14 $\pm$ 0.03. Preliminary results from a
survey of solar 
neighborhood K stars yields a very similar binary distribution
(Mayor et al. 1992), so the 
$csf_{5-50 AU,SN}$ should represent the $csf$ for nearby stars with
spectral types from F7 to K.

Recently, 
144 Pleiades G and K dwarfs were observed with adaptive optics at the CFHT
by Bouvier et al. (1997). These observations cover neither the same 
separation range nor the same range of sensitivity, complicating any
comparison between this data set and the Hyades speckle results. Due to the 
greater distance to the Pleiades, the minimum binary star separation observed 
in the Pleiades is 11 AU. In the Hyades survey presented here, 42\% of the 
speckle binaries have separations within the missing 5 to 11 AU range. 
Unlike the G-dwarf survey which has a comparable sensitivity to
the Hyades IR speckle survey, the Pleiades observations have a detection 
limit of at most $\Delta K = 2$ mag in the 11 to 50 AU range. 
The CFHT results have been corrected to allow for lower-mass companions
(to the Hydrogen-burning limit)
under the assumption that the companion star mass function is the same
as that of field stars (the procedure used to determine the Hyades 
$csf_{total,obs}$ in \S 6.2 gives the same correction for the Pleiades
as the one listed in the Bouvier et al. paper). 
This procedure results in very substantial 
corrections. Within the separation range overlapping the 
Hyades sample (11-50 AU), 7 binaries were observed, but an additional
12 undetected binaries are predicted. Including corrections for both the
missing separations and the undetectable companions,
the $csf_{5-50AU,Pleiades}$ is 0.23 $\pm$ 0.09. 
More sensitive observations are required before it is possible to make a
statistically significant comparison with the current Hyades data. 

Incorporating the results of the T Tauri, Hyades, and Solar Neighborhood surveys, 
Figure 5 shows the fraction of 
binaries with separations from 5 to 50 AU as a function of age. 
Because the comparison samples cover different mass and sensitivity ranges,
two values are computed for the Hyades sample. Since the T Tauri stars 
evolve into stars with masses $<3M_{\odot}$ and it is easier to detect
low mass companions when they are young (Ghez, White \& Simon 1997), the most appropriate Hyades $csf$ is
the entire main sequence sample (primary mass $\sim 0.6 M_{\odot}$ to $2.8 M_{\odot}$)
corrected to account for missing main sequence companions, $csf_{5-50AU,corr}$ of 0.30 $\pm$ 0.06. The solar neighborhood comparison is
more direct since the Duquennoy \& Mayor (1991) sample has the same sensitivity
level as the speckle observations and includes stars from F7 to G9 (with similar 
results for
K stars -- Mayor et al. 1992). The Hyades $csf_{5-50AU,obs}$ determined from the 
subset of 107 Hyades stars with (B-V) colors 
consistent with spectral types from F7 to K5 is most analogous to the 
solar neighborhood $csf$ and equals 0.21 $\pm$ 0.04. 
Although the
statistical significance of the differences are not high ($< 2\sigma$), the Hyades 
$csf$ is between the 
younger and older samples and may suggest a downward trend in multiplicity.
Observations of clusters with ages between the Hyades and T Tauri stars which 
cover a similar separation and sensitivity range are required to clearly 
establish whether or not an evolutionary trend in the companion star fraction
exists.

{\centering \leavevmode
\epsfxsize=.45\columnwidth \epsfbox{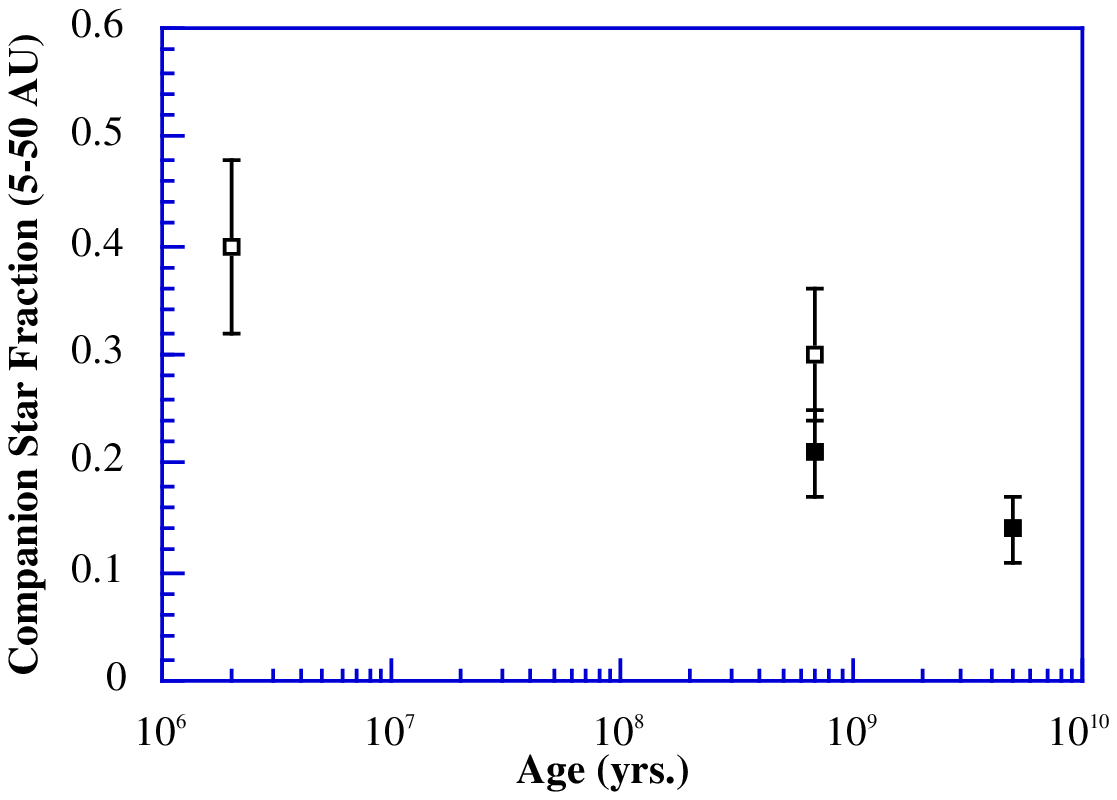}
\hspace*{.08\columnwidth}
\parbox[b]{.45\columnwidth}{Figure~5: The $csf_{5-50AU}$ for three stellar samples is plotted as a function of 
the sample age. Open squares signify the $csf$ of stars with masses from 
$\sim 0.5M_{\odot}$ to $\sim$3$M_{\odot}$, the full range of the sample,
which overlaps the mass range of T Tauri stars.
The Hyades value has been corrected to account for all stellar 
companions, and the corrected T Tauri value has a similar sensitivity. 
Filled squares represent the }} 
\noindent
$csf_{5-50AU}$ of late-F through K stars,
the spectral type range covered by the Duquennoy \& Mayor (1991) solar 
neighborhood survey.
The Hyades value is not corrected since the Hyades survey has a detection
limit comparable to that reported for the solar neighborhood results.
The figure is suggestive of an evolutionary trend in 
multiplicity, although the $csf_{5-50AU}$ measured for the Hyades is not significantly
different from the $csf_{5-50AU}$ of the older solar neighborhood.

\subsection{Mass Ratio Distribution -- an Observational Test of 
Several Binary Star Formation Mechanisms}

The mass ratio (q) distribution and its dependence on separation, primary mass, 
and radial distance provide 
additional constraints on several binary star formation theories. 
The mass ratio distribution for all binaries with 
separations from 5 to 50 AU is shown in Figure 6 and increases toward
smaller mass ratios from 1.00 down to a ratio of 0.30. The decrease in the distribution for
mass ratios below 0.3 is due to incompleteness. Only half of the observations
are sensitive to mass ratios of 0.23, whereas {\sl all} the observations are
sensitive to mass ratios greater than 0.30. 
To avoid any observational bias, this analysis is restricted to mass 
ratios from 0.30 to 1.00. The best fit power law description of the data
is $q^{-1.3 \pm 0.3}$ which has a K-S test probability of 
89\%. 

{\centering \leavevmode
\epsfxsize=.45\columnwidth \epsfbox{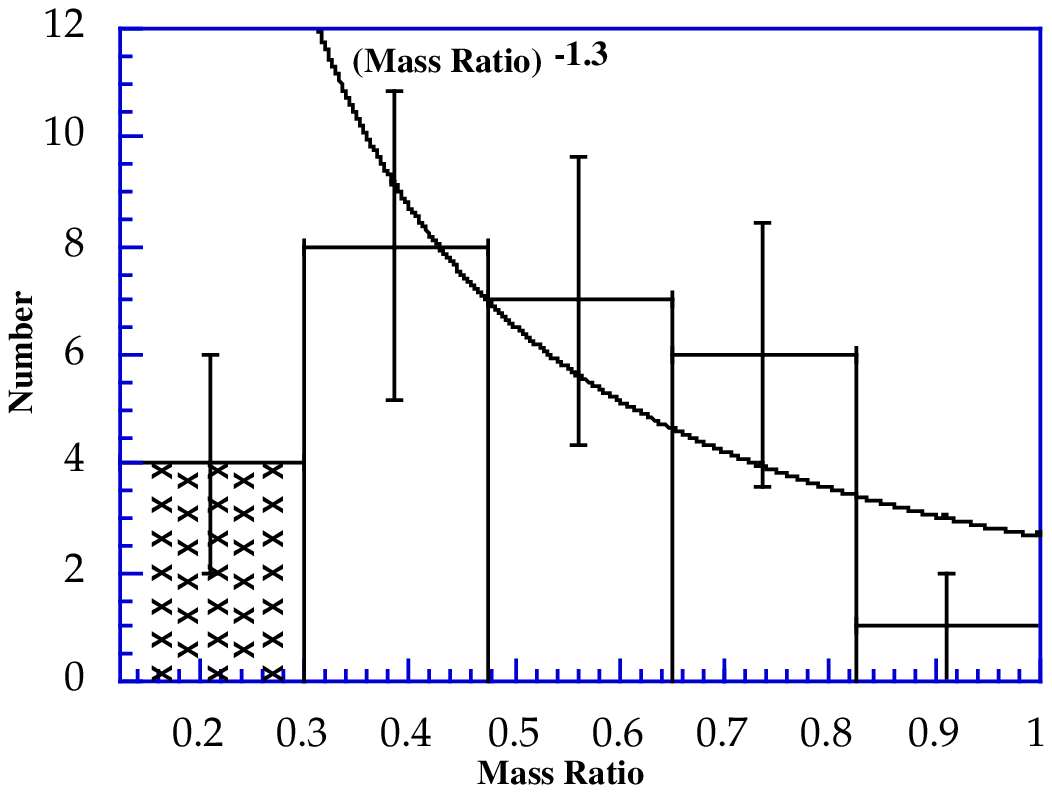}
\hspace*{.08\columnwidth}
\parbox[b]{.45\columnwidth}{Figure~6: A histogram of mass ratios, $q$, for the binaries with separations from 0\farcs10 to
1\farcs07 is shown. Since all the observations in the survey are 
sensitive enough
to detect a binary with a mass ratio of 0.30, only the 22 binaries with
mass ratios larger than 0.30 were involved in determining the fit to the 
mass ratio distribution. The additional 4 systems with smaller mass ratios
are included in }}
\noindent
the graph, but the bin containing these binaries is marked
with x's since the survey is not complete at this mass ratio extreme. The
best fit power law of $q^{-1.3}$ is also shown.

This declining power law is inconsistent with the flat or 
slightly rising distributions of both 
30 Pleiades F7-K0 spectroscopic and photometric binaries and 23 solar 
neighborhood G-dwarf
spectroscopic systems (Mermilliod et al. 1992, Mazeh et al. 1992). 
The distribution of the 23 G-dwarf binaries with periods less than 3000
days was found to be different from the long period distribution of G-dwarf
binaries. Dividing the complete sample of Hyades speckle binaries in half
based on separation showed no evidence of a separation dependence for the 
mass ratio distribution; the K-S probability that the two separation 
distribution are the same 
is 89\%. Because the separation range of the speckle observations is limited
to 45 AU, comparison with spectroscopic or visual binaries may be required
to test for a difference in the mass ratio distributions. This comparison,
however, has the advantage of studying distributions of mass ratios constructed
with binaries detected by the same technique.

The mass ratio distribution also shows no dependence on either primary 
star mass or radial distance. To 
investigate the mass dependence of the distribution, the binaries 
are divided in half based on their 
primary mass and a K-S test indicates that the mass ratio distributions for 
high and low mass primaries have an 81\% probability of being the same;
there is no mass dependence of the mass ratio distribution. Similarly, the
mass ratio distribution does not depend on radial distance; the half of the
binaries with smaller radial distances has an 81\% probability of being drawn
from the same distribution as the half of the binaries with larger radial
distances.

Both the scale-free fragmentation model and capture
in small-N clusters
make specific predictions that can be compared to the observational results
described above. 
For binaries formed by scale-free fragmentation, the mass 
ratio distribution is expected to be independent of, or only weakly dependent on, the primary star mass
(Clarke 1998), consistent with the Hyades data. This formation scenario
is also consistent with the observation that the $csf$ is independent of mass.
Diskless capture in small-N clusters tends to form binaries consisting of
the two most massive stars (McDonald \& Clarke 1995), causing the distribution
to increase towards large mass ratios; with its negative slope, the Hyades
mass ratio distribution, like the $csf$ in \S 6.1.2, does not support this 
model. 

Simulations of accretion during binary formation also 
predict measurable effects in the mass ratio distribution.
This model suggests that accretion of 
high angular momentum circumbinary material drives the mass ratio toward unity,
while accretion of low angular momentum circumstellar material results in 
smaller mass ratios (Bate \& Bonnell 1997). The Hyades data suggest that few
binaries have accreted a large amount of circumbinary material. 
A number of scenarios such as scale dependent fragmentation and disk 
fragmentation (cf. Boss \& Myhill 1995, Myhill \& 
Kaula 1992, Burkert \& Bodenheimer 1996, Bonnell \& Bate 1994)
which lack observational predictions remain possible formation
mechanisms in addition to the scale-free fragmentation model.

\section{Summary}
Infrared speckle observations of 167 bright Hyades members, approximately 
one-third of the cluster, were made with the Hale 5m telescope. A total of
33 binaries were resolved, of which 13 are new detections, and an additional
17 are known spectroscopic binaries. Including the results from spectroscopic
and direct imaging surveys, the ratio of singles:binaries:triples in the 
sample is 98:59:10.

Over the
separation range $0\farcs10$ to $1\farcs07$, the observations are sensitive
to companions 4 mag fainter than the target star. Within this 
separation
range, 26 of the 162 main sequence stars are resolved as binaries, resulting
in an observed $csf_{5-50AU,corr}$ of 0.16 $\pm$ 0.03; accounting 
for the inability to 
detect fainter companions increases the multiplicity to 
$csf_{5-50AU,total}$ of
0.30 $\pm$ 0.06. The 
Hyades $csf$ is intermediate between the fractions of the younger T Tauri 
stars and the older solar neighborhood. Although the observations permit an 
evolutionary trend in multiplicity, this result is not conclusive and future
observations of other young clusters will further illuminate this discussion.

Within the Hyades speckle sample, the $csf$ is independent of radial distance 
and primary star mass. Another key observational result is the mass ratio 
distribution. Unlike spectroscopic studies which are
biased because of the uncertainty of the inclination angle, the resolved 
speckle binaries provide mass ratios that are free of selection effects 
from ratios of 0.30 to 1.00. The observed mass ratio distribution is best 
described by a power law $q^{-1.3 \pm 0.3}$. This mass ratio distribution 
does not vary with primary star mass, binary star separation or 
distance from the cluster center.
Comparing  models of accretion during binary formation to the observed mass
ratio distribution leads to 
the conclusion that few binaries experience accretion of high angular momentum
material. Overall, the Hyades data support the scale-free fragmentation model, but not 
capture in small-N clusters or disk-assisted capture in small-N clusters
(McDonald \& Clarke 1993, 1995; Clarke 1998). 
In addition to scale-free fragmentation, binary star
formation mechanisms not rejected by the Hyades data are scale dependent 
fragmentation and disk fragmentation, scenarios
for which there are currently no observational tests (cf. Boss \& 
Myhill 1995, Myhill \& 
Kaula 1992, Burkert \& Bodenheimer 1996, Bonnell \& Bate 1994).

\acknowledgments

Support for this work was provided by NASA through grant number
NAGW-4770 under the Origins of Solar Systems Program. We thank the 
staff at Palomar for their assistance with our observations. This 
research required extensive use of the SIMBAD database, operated at
CDS, Strasbourg, France.

\appendix

\section{An Empirical Mass-$M_K$ Relation}

The results and limitations of this survey are transformed into physical 
parameters through an empirical mass-$M_K$ relation. 
Since this relation varies with age and metallicity, 
the ideal relation
would be constructed from Hyades stars. Although the nearby star samples
do not have the same age and metallicity as the Hyades, the masses of a 
number of these stars have been determined (Andersen 1991, Henry \& McCarthy
1993), and the empirical relation used
for the Hyades stars is based on solar neighborhood surveys.  Because 
many of the stars in the Hyades sample have $M_K <$ 3.07,
the relations derived by Henry \& McCarthy (1993) cannot be applied to the 
entire sample. 
An alternate mass-$M_K$ relation was constructed by combining the
low mass Henry \& McCarthy data with the higher mass data 
listed for Main Sequence detached eclipsing binaries in the review 
by Andersen. The $M_V$ given for each star in the more massive systems 
was converted into an $M_K$ based on a color-color
relation constructed with the data compiled in Kenyon \& Hartmann (1995). The 
linear fit to the B-V and V-K data listed for A through K stars is

\begin{equation}
(V-K) = 2.38(B-V) + 0.03.
\end{equation}

\noindent
A single line was used to fit the A through M star data rather than
a combination of three lines as in Henry \& McCarthy (1992), and the 
resulting mass-$M_K$ relation is 

\begin{equation}
log(M/M_{\odot}) = -0.159M_K + 0.49.
\end{equation}

\noindent
This relation is used to convert each observed binary $\Delta K$ or 
single $\Delta K_{lim}$ into a mass ratio or a mass ratio detection limit.
For fainter magnitudes, the fit predicts that the Hydrogen-burning limit of 
0.08$M_{\odot}$ occurs at $M_K \sim 10$. For brighter magnitudes,
the recently-determined dynamical masses for the components of vB 24 and vB 57
provide a check on the empirical relation at higher masses (Torres et al. 
1997a,b). For both binaries,
the photometric masses derived from the infrared speckle measurements match 
the dynamical values
almost exactly for the primary mass. The average discrepancy in the 
secondary mass 
is 23\%, and this value is taken as the uncertainty in the measurements of 
the secondary masses and the mass ratios.

An empirical mass - $M_V$ relation, also 
constructed from the same data set, 
is necessary to compare the infrared observations presented here with the
previous work done at optical wavelengths. The mass-$M_V$ relation is

\begin{equation}
log(M/M_{\odot}) = -0.090M_V + 0.45.
\end{equation}

\noindent
 Because the mass-$M_V$ relation has a 
shallower slope than the corresponding mass-$M_K$ relation, a larger 
$\Delta V$ than a $\Delta K$ detection limit is required to reach the same 
companion mass. Due to this effect, an optical speckle survey with similar
dynamic range as the infrared speckle observations, is much less sensitive
to low mass companions. With a detection limit of $\Delta V = 3$, the optical 
speckle survey conducted by Mason et al. (1993) has a mass ratio limit of 0.54.
The visual pair binaries in the Duquennoy \& Mayor (1991) survey were
corrected to a larger value of $\Delta V$, 7 mag, which corresponds to a mass 
ratio limit of 0.23, similar to the Hyades survey. Both the G-dwarf and the
Hyades surveys are sensitive to companions as faint as early M stars.

\end{document}


\tiny
\begin{deluxetable}{lllllcccll}
\tablenum{3}
\tablecaption{Hyades Speckle Binaries}
\tablewidth{7.5 in}
\tablehead{ \colhead{Object}
& \colhead{Date}
& \colhead{$\Delta K$}
& \colhead{Separation}
& \colhead{Position Angle}
& \colhead{$M_1$}
& \colhead{$M_2$}
& \colhead{Comp.}
& \multicolumn{2}{c}{Notes on$^d$} \nl
\colhead{}
& \colhead{(UT)}
& \colhead{(mag)}
& \colhead{(arcsec)}
& \colhead{(deg)}
& \colhead{($M_{\odot}$)}
& \colhead{($M_{\odot}$)}
& \colhead{}
& \colhead{Binaries}
& \colhead{Triples}}
\startdata

vB 96 & 1995 Nov 13 & 0.70 $\pm$ 0.02$^c$ & 0.048 $\pm$ 0.001 & 288 $\pm$ 4 & 0.85 & 0.66 & 1 & M: $0\farcs192$, G: $\sim$5000dy\nl
        & 1996 Dec 23$^b$ & 0.70 $\pm$ 0.02 & 0.092 $\pm$ 0.002 & 330 $\pm$ 1  \nl 
+10 568 & 1996 Jan 09 & 0.14 $\pm$ 0.02$^c$ & 0.043 $\pm$ 0.001 & 213 $\pm$ 1 & 0.59 & 0.56 & 1 & new \nl
        & 1997 Dec 14$^b$ & 0.14 $\pm$ 0.2 & 0.0463 $\pm$ 0.001 & 154 $\pm$ 1 &  &  &  & \nl
vB 91 & 1994 Dec 22 & 1.02 $\pm$ 0.02$^c$ &  0.058 $\pm$ 0.001  &  339 $\pm$ 10 & 0.81 & 0.56 & 1 & M: $0\farcs192$, G: $>$6000dy\nl
 & 1996 Dec 22$^b$ & 1.02 $\pm$ 0.02 & 0.115 $\pm$ 0.002 & 340 $\pm$ 1 & & \nl
vB 120 & 1996 Jan 10 & 0.25 $\pm$ 0.03 & 0.069 $\pm$ 0.002 & 31 $\pm$ 1 & 1.05 & 0.96 & 1 & M: $0\farcs082$, B\&S: ?\nl
Lei 90 & 1994 Dec 22 & 2.35 $\pm$  0.07 & 0.084 $\pm$ 0.003 & 12 $\pm$ 3 & 0.75 & 0.32 & 1 & G: 3942dy\nl
vB 57 & 1993 Nov 28 & 0.58 $\pm$ 0.03 & 0.086 $\pm$ 0.002 & 289 $\pm$ 1 & 1.32 & 1.07 & 1 & M: $0\farcs099$, G: 2192dy, IDS: $0\farcs1$\nl
  & & & & \nl
vB 113 & 1995 Nov 15 & 2.7 $\pm$ 0.4 &  0.098 $\pm$ 0.002 & 124 $\pm$ 1 & 1.02 & 0.38 & 1 & G: 2557dy\nl
vB 81 & 1996 Jan 10 & 2.13 $\pm$ 0.06 & 0.107 $\pm$ 0.002 & 114 $\pm$ 5 & 1.20 & 0.55 &  1 & B\&S: ?\nl
vB 24 & 1993 Nov 30 & 2.3 $\pm$ 0.4 &   0.139 $\pm$  0.008 & 166 $\pm$ 6 & 1.80 & 1.00 & 1 & M: $0\farcs096$, S\&L: 4177dy\nl
 & 1996 Jan 10 & 1.6 $\pm$ 0.1 & 0.146 $\pm$ 0.003 & 192 $\pm$ 2 &  & \nl
Lei 59 &  1996 Jan 09 & 0.64 $\pm$ .03 & 0.137 $\pm$ 0.003 & 65 $\pm$ 1 & 0.69 & 0.54 & 1 & new \nl
vB 103 & 1996 Jan 09 & 2.5 $\pm$ 0.2 &  0.158 $\pm$ 0.008 & 220 $\pm$ 1 & 1.70 & 0.68 & 1 & new \nl
vB 114 & 1994 Dec 21  & 3.9 $\pm$ 0.1 & 0.17 $\pm$ 0.02 & 108 $\pm$ 3 & 0.87 & 0.21 & 1 & G: $>$1826dy\nl
vB 50 & 1993 Nov 28 & 1.39 $\pm$  0.06 & 0.171 $\pm$ 0.005 & 180 $\pm$ 1 & 0.95 & 0.57 & 1 & M: $0\farcs262$, B\&S: ?\nl
vB 59 & 1994 Dec 21 & 3.6 $\pm$ 0.2 & 0.184 $\pm$  0.006 & 79 $\pm$ 2 & 1.07 & 0.28 & 1 & G: $>$2557dy\nl
vB 58 & 1994 Dec 21 & 0.60 $\pm$ 0.01 & 0.185 $\pm$ 0.004 & 162 $\pm$ 1 & 0.97 & 0.78 & 1 & S\&L: 10077dy, ADS: $0\farcs34$\nl 
Lei 20 & 1995 Nov 13 & 1.04 $\pm$ 0.04 & 0.202 $\pm$ 0.005 & 191 $\pm$ 1 & 0.72 & 0.49 & 2 & new & G\&G81: 2dy\nl
Lei 83 & 1994 Dec 22 & 2.2 $\pm$ 0.04 & 0.212 $\pm$ 0.006 & 98 $\pm$ 3 & 0.68 & 0.30 & 2 & new & G85: 61dy\nl
vB 102 & 1996 Jan 09 & 2.0 $\pm$ 0.1 & 0.232 $\pm$  0.005 & 280 $\pm$ 1 & 1.00 & 0.49 & 2 & M: $0\farcs235$ &  G: 731dy\nl
vB 85 & 1995 Nov 13 & 3.0 $\pm$ 0.2 &  0.252 $\pm$ 0.006 & 63 $\pm$  1 & 1.39 & 0.46 & 1 & new \nl
vB 122 & 1995 Nov 15 & 0.0 $\pm$ 0.1 & 0.254 $\pm$ 0.006 & 142 $\pm$ 1 & 1.13 & 1.13 & 1 & M: $0\farcs209$, ADS: $0\farcs13$, S\&L: 5947dy \nl
vB 75 & 1993 Nov 28 & 0.9 $\pm$  0.2 & 0.305 $\pm$ 0.008 & 103 $\pm$ 1 & 1.49 & 1.09 & 1 & M: $0\farcs411$, S\&L: 14584dy, ADS: $0\farcs11$ \nl
Lei 92 & 1994 Dec 22 & 1.72 $\pm$ 0.01 & 0.310 $\pm$ 0.006 & 87 $\pm$ 1 & 0.69 & 0.37 & 1 & new \nl
vB 29 & 1995 Nov 14 & 0.90 $\pm$ 0.02 & 0.321 $\pm$ 0.007 &  75 $\pm$ 1 & 1.17 & 0.84 & 1 & ADS: $0\farcs33$, S\&L: 32595dy \nl
vB 185 & 1994 Dec 22 & 1.11 $\pm$ 0.01 & 0.443 $\pm$ 0.009 & 359 $\pm$ 1 & 0.80 & 0.53 & 2 & M: $0\farcs659$ & G85: 277dy \nl
vB 124 & 1995 Nov 15 & 1.3 $\pm$ 0.2 & 0.46 $\pm$ 0.01 & 153 $\pm$ 1 & 1.61 & 1.01 & 2 & M: $0\farcs452$, ADS: $0\farcs1$, S\&L: 34786dy & G85: 143dy\nl
vB 17 & 1994 Dec 21 & 3.9 $\pm$ 0.3 & 0.73 $\pm$  0.02 & 355 $\pm$ 1 & 0.85 & 0.21 & 1 & new \nl
vB 151 & 1996 Jan 09 & 1.45 $\pm$ 0.06 & 0.85 $\pm$ 0.02 & 312 $\pm$ 1 & 0.64 & 0.37 & 2 & new & G: 731dy \nl
vB 5$^a$ & 1994 Dec 22 & 2.31 $\pm$ 0.05 &   0.88 $\pm$ 0.02 & 85 $\pm$ 1 & 0.68 & 0.29 & 1 & new \nl
Lei 52$^a$ & 1995 Nov 13 & 5.5 $\pm$ 0.4 & 0.91 $\pm$ 0.02 & 15 $\pm$ 1 & 0.76 & 0.10 & 1 & new \nl
Lei 130$^a$ & 1995 Nov 15 & 1.2 $\pm$ 0.1 & 0.99 $\pm$ 0.02 & 128 $\pm$ 1 & 0.58 & 0.38 & 1 & new \nl
vB 52$^a$ & 1996 Jan 10 & 2.50 $\pm$ 0.07 & 1.02 $\pm$ 0.02 & 149 $\pm$ 1 & 0.91 & 0.36 & 1 & B\&S: ?\nl
+22 669$^a$ & 1995 Nov 13 & 2.9 $\pm$ 0.1 & 1.02 $\pm$ 0.02 & 208 $\pm$ 2 & 0.89 & 0.31 & 2 & new & G82: 2dy \nl 
   & & & & \nl
vB 40$^a$ & 1993 Nov 28 & 2.9 $\pm$ 0.5 &  1.33 $\pm$  0.04 & 193 $\pm$ 1 & 1.15 & 0.39 & 2 & ADS: $1\farcs16$  &  G,S: 4dy \nl

\tablecomments{$^a$ - Binary parameters determined from shift and add analysis \nl
$^b$ - Keck measurement, $^c$ - Fixed to Keck flux ratio value \nl
$^d$ - M - Mason et al. 1993, G - Griffin et al. 1988, B\&S - Barrado y Navascues \& Stauffer 1997, period unlisted, counted as binary not triple (see text) \nl
IDS - IDS catalogue, S\&L - Stefanik \& Latham 1992, ADS - ADS catalogue, G\&G81 - Griffin \& Gunn 1981, G85 - Griffin et al. 1985 \nl
G82 - Griffin et al. 1982, G,S - Sanford 1921, listed in Griffin et al. 1988 }

\enddata
\end{deluxetable}


\small
\begin{deluxetable}{lcccccl}
\tablenum{4}
\tablecaption{Limits for Undetected Companions to Hyades Speckle Singles}
\tablewidth{6.2 in}
\tablehead{ \colhead{Object}
& \colhead{Date (UT)}
& \colhead{$\Delta K_{lim}$}
& \colhead{$M (M_{\odot})$}
& \colhead{$M_{lim} (M_{\odot}$)}
& \colhead{$\#$ Comp.}
& \colhead{Notes on Other Measurements$^a$}}
\startdata

+20 480  &  1996 Jan 09  & 3.69 & 1.36 & 0.35 & 0 &  \nl
vB 1  &  1995 Nov 14  & 4.46 & 1.05 & 0.20 & 0 &  \nl
Lei 2  &  1995 Nov 14  & 4.29 & 0.62 & 0.13 & 0 &  \nl
+35 714  &  1994 Dec 21  & 4.48 & 1.19 & 0.23 & 1 & M: $0\farcs222$  \nl
vB 4  &  1994 Dec 22  & 3.76 & 0.78 & 0.19 & 0 &  \nl
+16 516  &  1995 Nov 13  & 3.20 & 0.83 & 0.26 & 1 & N\&Y: 0.5dy(WD) \nl
vB 170  &  1995 Nov 15  & 4.03 & 0.61 & 0.14 & 0 & \nl 
vB 6  &  1993 Nov 28  & 3.73 & 1.49 & 0.38 & 0 &  \nl
vB 7  &  1995 Nov 13  & 3.86 & 0.79 & 0.19 & 0 &  \nl
vB 8  &  1996 Jan 10  & 2.87 & 1.36 & 0.47 & 0 &  \nl
Lei 11  &  1995 Nov 13  & 3.25 & 0.63 & 0.19 & 0 & \nl 
Lei 10  &  1994 Dec 22  & 3.54 & 0.80 & 0.22 & 0 &  \nl
vB 10  &  1996 Jan 10  & 4.16 & 0.99 & 0.22 & 0 &  \nl
Lei 15  &  1996 Jan 10  & 4.50 & 0.66 & 0.13 & 0 &  \nl
Lei 16  &  1995 Nov 13  & 3.54 & 0.64 & 0.17 & 0 &  \nl
vB 11  &  1993 Nov 28  & 3.01 & 1.47 & 0.49 & 1 & ADS: $2\farcs0$ \nl
+13 647  &  1994 Dec 21  & 3.98 & 1.12 & 0.26 & 0 &  \nl
Lei 18  &  1995 Nov 13  & 3.66 & 0.80 & 0.21 & 0 & \nl 
+8 642  &  1995 Nov 15  & 3.86 & 0.72 & 0.17 & 0 &  \nl
vB 13  &  1994 Dec 21  & 4.76 & 1.32 & 0.23 & 0 &  \nl
vB 14  &  1996 Jan 10  & 3.25 & 1.47 & 0.45 & 1 & B\&S: ?  \nl
vB 16  &  1996 Jan 10  & 3.73 & 1.36 & 0.35 & 0 &  \nl
vB 15  &  1994 Dec 21  & 3.62 & 0.91 & 0.24 & 0 &  \nl
vB 18  &  1994 Dec 21  & 4.08 & 0.96 & 0.22 & 0 &  \nl
vB 19  &  1995 Nov 15  & 3.80 & 1.18 & 0.29 & 0 &  \nl
vB 162  &  1996 Jan 10  & 2.87 & 1.12 & 0.39 & 1 & G\&G81: 55dy \nl
vB 20  &  1995 Nov 15  & 4.03 & 1.48 & 0.34 & 0 &  \nl
vB 21  &  1995 Nov 13  & 3.62 & 0.78 & 0.21 & 0 &  \nl
vB 22  &  1994 Dec 21  & 4.43 & 1.01 & 0.20 & 1 & G85: 6dy \nl
vB 23  &  1996 Jan 10  & 4.01 & 1.23 & 0.28 & 1 & B\&S: ?  \nl
vB 25  &  1994 Dec 21  & 4.25 & 0.70 & 0.15 & 0 &  \nl
vB 26  &  1996 Jan 10  & 4.11 & 0.85 & 0.19 & 0 &  \nl
vB 27  &  1995 Nov 15  & 3.73 & 0.93 & 0.24 & 0 &  \nl
vB 28  &  1996 Jan 09  & 1.04 & giant & 4.38 & 0 &  \nl
vB 30  &  1995 Nov 14  & 3.25 & 1.66 & 0.50 & 0 &  \nl
vB 31  &  1995 Nov 15  & 4.35 & 1.09 & 0.22 & 0 &  \nl
vB 32  &  1996 Jan 10  & 3.08 & 1.51 & 0.49 & 0 &  \nl
vB 33  &  1996 Jan 10  & 3.73 & 1.89 & 0.48 & 0 &  \nl
vB 34  &  1993 Nov 28  & 3.89 & 1.70 & 0.41 & 2 & S\&L: 3dy, BV: ?(WD)  \nl
vB 35  &  1993 Nov 28  & 4.08 & 1.34 & 0.30 & 0 &  \nl
vB 36  &  1993 Nov 28  & 3.01 & 1.29 & 0.43 & 0 &  \nl
vB 37  &  1994 Dec 21  & 4.48 & 1.36 & 0.26 & 0 &  \nl
vB 38  &  1995 Nov 14  & 3.58 & 1.68 & 0.45 & 1 & S\&L: 2dy \nl
vB 39  &  1996 Jan 10  & 2.70 & 0.89 & 0.33 & 1 & G: $>$2557dy  \nl
vB 41  &  1996 Jan 09  & 4.58 & giant & 1.19 & 2 & M: $0\farcs273$, G: 530dy \nl
vB 42  &  1996 Jan 09  & 4.53 & 0.85 & 0.16 & 0 &  \nl
vB 43  &  1995 Nov 15  & 4.03 & 0.80 & 0.18 & 1 & G85: 591dy  \nl
vB 44  &  1996 Jan 10  & 2.94 & 1.26 & 0.43 & 0 &  \nl
vB 45  &  1995 Nov 14  & 3.80 & 1.78 & 0.44 & 1 & S\&L: 8dy \nl
vB 46  &  1994 Dec 21  & 4.20 & 0.75 & 0.16 & 0 &  \nl
vB 47  &  1996 Jan 09  & 4.20 & 2.09 & 0.45 & 0 &  \nl
vB 48  &  1993 Nov 28  & 3.66 & 1.13 & 0.30 & 0 &  \nl
vB 49  &  1995 Nov 15  & 3.40 & 0.92 & 0.26 & 0 &  \nl
vB 174  &  1995 Nov 15  & 4.20 & 0.73 & 0.16 & 0 &  \nl
vB 51  &  1993 Nov 28  & 3.98 & 1.27 & 0.30 & 0 &  \nl
vB 53  &  1995 Nov 14  & 3.95 & 1.65 & 0.39 & 0 &  \nl
vB 140  &  1996 Jan 09  & 4.33 & 0.81 & 0.17 & 1 & G85: 156dy \nl
vB 175  &  1995 Nov 15  & 4.03 & 0.70 & 0.16 & 0 &  \nl
vB 54  &  1994 Dec 23  & 4.52 & 2.61 & 0.50 & 0 &  \nl
vB 55  &  1996 Jan 09  & 4.31 & 1.86 & 0.38 & 0 &  \nl
vB 56  &  1996 Jan 09  & 4.31 & 2.23 & 0.46 & 0 &  \nl
vB 60  &  1994 Dec 23  & 4.23 & 2.76 & 0.59 & 0 &  \nl
vB 62  &  1995 Nov 14  & 3.54 & 1.18 & 0.32 & 1 & G\&G78: 9dy \nl
vB 141  &  1996 Jan 09  & 3.66 & 2.44 & 0.64 & 1 & A: 5200dy \nl
vB 68  &  1996 Jan 10  & 4.18 & 1.61 & 0.35 & 0 &  \nl
vB 63  &  1995 Nov 14  & 2.94 & 0.95 & 0.32 & 1 & G: 2557dy \nl
vB 64  &  1996 Jan 09  & 4.03 & 0.94 & 0.21 & 0 &  \nl
Lei 49  &  1995 Nov 13  & 3.49 & 0.82 & 0.66 & 0 &  \nl
vB 177  &  1994 Dec 21  & 4.31 & 0.62 & 0.13 & 0 &  \nl
vB 65  &  1995 Nov 13  & 4.27 & 1.08 & 0.23 & 0 &  \nl
vB 66  &  1995 Nov 14  & 4.48 & 1.09 & 0.21 & 0 &  \nl
vB 67  &  1995 Nov 15  & 3.98 & 1.73 & 0.40 & 0 &  \nl
Lei 57  &  1994 Dec 22  & 4.31 & 0.68 & 0.14 & 1 & G85: 1907dy  \nl
Lei 50  &  1994 Dec 22  & 4.01 & 0.86 & 0.20 & 0 &  \nl
vB 69  &  1995 Nov 14  & 3.01 & 0.90 & 0.30 & 1 & G85: 42dy \nl
vB 70  &  1996 Jan 09  & 4.5 & giant & 1.33 & 0 &  \nl
vB 71  &  1996 Jan 09  & 4.63 & giant & 1.07 & 1 & M: $0\farcs048$, G: 5844dy  \nl
vB 72  &  1994 Dec 23  & 4.01 & evolved & 0.80 & 1 & S\&L: 140dy \nl
vB 73  &  1995 Nov 14  & 3.76 & 0.95 & 0.24 & 0 &  \nl
vB 74  &  1994 Dec 23  & 5.10 & 2.04 & 0.31 & 0 &  \nl
vB 76  &  1996 Jan 10  & 2.94 & 1.73 & 0.39 & 0 &  \nl
vB 77  &  1995 Nov 14  & 3.31 & 0.80 & 0.27 & 1 & G85: 239dy  \nl
vB 78  &  1996 Jan 10  & 3.01 & 1.22 & 0.40 & 0 &  \nl
vB 79  &  1994 Dec 22  & 3.62 & 0.80 & 0.21 & 0 &  \nl
Lei 55  &  1994 Dec 22  & 3.76 & 0.77 & 0.19 & 0 &  \nl
Lei 56  &  1994 Dec 22  & 4.13 & 0.81 & 0.18 & 0 &  \nl
vB 82  &  1994 Dec 23  & 4.76 & 2.12 & 0.37 & 0 &  \nl
vB 182  &  1994 Dec 22  & 3.95 & 0.90 & 0.21 & 1 & G\&G81: 358dy \nl
vB 83  &  1995 Nov 13  & 4.13 & 1.78 & 0.39 & 1 & A\&L: 106.3dy \nl
vB 84  &  1995 Nov 13  & 3.83 & 1.74 & 0.43 & 0 &  \nl
vB 86  &  1996 Jan 09  & 4.53 & 1.32 & 0.25 & 0 &  \nl
vB 87  &  1996 Jan 09  & 4.55 & 0.89 & 0.17 & 0 &  \nl
Lei 63  &  1994 Dec 22  & 3.76 & 0.71 & 0.18 & 1 & G: 845dy \nl
vB 89  &  1995 Nov 14  & 3.95 & 1.54 & 0.36 & 0 &  \nl
vB 90  &  1996 Jan 09  & 4.31 & 1.33 & 0.27 & 0 &  \nl
vB 92  &  1994 Dec 22  & 4.31 & 0.92 & 0.19 & 0 &  \nl
vB 93  &  1994 Dec 22  & 4.11 & 0.86 & 0.19 & 0 &  \nl
vB 94  &  1995 Nov 13  & 3.95 & 1.44 & 0.34 & 0 &  \nl
vB 95  &  1994 Dec 23  & 4.53 & 2.35 & 0.45 & 1 & A: 488.5dy \nl
vB 183  &  1994 Dec 22  & 3.80 & 0.82 & 0.20 & 0 &  \nl
vB 97  &  1995 Nov 13  & 4.45 & 1.07 & 0.21 & 0 &  \nl
vB 99  &  1994 Dec 22  & 3.73 & 0.81 & 0.21 & 0 &  \nl
vB 100  &  1995 Nov 15  & 3.20 & 1.56 & 0.48 & 0 &  \nl
vB 101  &  1995 Nov 14  & 3.49 & 1.48 & 0.41 & 0 &  \nl
vB 210  &  1995 Nov 14  & 4.03 & 0.73 & 0.17 & 0 &  \nl
vB 104  &  1994 Dec 23  & 4.82 & 2.38 & 0.4 & 0 &  \nl
vB 105  &  1995 Nov 15  & 4.33 & 1.04 & 0.21 & 0 &  \nl
vB 106  &  1995 Nov 14  & 4.20 & 1.02 & 0.22 & 1 & G: 3653dy \nl
vB 107  &  1996 Jan 09  & 4.23 & 1.88 & 0.40 & 0 &  \nl
vB 108  &  1994 Dec 23  & 4.52 & 2.47 & 0.47 & 0 &  \nl
+12 623  &  1994 Dec 22  & 4.08 & 0.67 & 0.15 & 0 &  \nl
vB 109  &  1995 Nov 15  & 3.69 & 0.93 & 0.24 & 0 &  \nl
vB 111  &  1996 Jan 09  & 4.39 & 1.76 & 0.35 & 0 &  \nl
vB 112  &  1995 Nov 15  & 3.58 & 2.23 & 0.60 & 1 & A\&L: 18dy \nl
vB 142  &  1994 Dec 22  & 4.60 & 0.93 & 0.17 & 1 & B\&S: ?  \nl
vB 115  &  1994 Dec 21  & 4.60 & 0.82 & 0.15 & 1 & G: 1461dy \nl
vB 116  &  1996 Jan 10  & 3.08 & 0.78 & 0.25 & 0 &  \nl
vB 117  &  1995 Nov 15  & 4.35 & 0.78 & 0.16 & 1 & G\&G78: 12dy \nl
vB 118  &  1994 Dec 22  & 3.89 & 1.06 & 0.25 & 0 &  \nl
vB 119  &  1995 Nov 13  & 3.49 & 1.18 & 0.33 & 1 & B\&S: ? \nl
vB 121  &  1995 Nov 13  & 4.48 & 1.20 & 0.23 & 1 & G\&G78: 6dy \nl
vB 123  &  1995 Nov 14  & 3.76 & 2.10 & 0.53 & 0 & \nl 
vB 143  &  1994 Dec 22  & 4.20 & 1.21 & 0.26 & 0 &  \nl
Lei 98  &  1995 Nov 15  & 3.08 & 0.75 & 0.24 & 0 &  \nl
vB 126  &  1995 Nov 14  & 3.36 & 1.56 & 0.46 & 0 &  \nl
vB 127  &  1994 Dec 22  & 3.92 & 0.94 & 0.22 & 0 &  \nl
vB 128  &  1996 Jan 09  & 4.01 & 1.27 & 0.29 & 0 &  \nl
vB 129  &  1995 Nov 14  & 3.69 & 2.51 & 0.65 & 0 &  \nl
Lei 107  &  1995 Nov 15  & 3.36 & 0.91 & 0.27 & 0 &  \nl
vB 215  &  1995 Nov 15  & 3.80 & 0.97 & 0.24 & 0 &  \nl
vB 202  &  1996 Jan 10  & 3.08 & 0.69 & 0.22 & 0 &  \nl
vB 130  &  1996 Jan 09  & 4.08 & 2.01 & 0.45 & 1 & B\&C: 155.8dy \nl
vB 131  &  1996 Jan 10  & 3.92 & 1.76 & 0.42 & 1 & S\&L: 11725dy \nl
vB 132  &  1994 Dec 21  & 3.08 & 1.13 & 0.37 & 1 & M: $0\farcs290$, ADS: $0\farcs31$ \nl
 
\tablecomments{$^a$ - M - Mason et al. 1993, N\&Y - Nelson \& Young 1976, ADS - ADS catalogue \nl 
B\&S - Barrado y Navascues \& Stauffer 1997 (period unlisted), G\&G81 - Griffin \& Gunn 1981 \nl
G85 - Griffin et al. 1985, S\&L - Stefanik \& Latham 1992, BV - Bohm-Vitense 1993, G - Griffin et al. 1988 \nl
G\&G78 - Griffin \& Gunn 1978, A - Abt 1965, A\&L - Abt \& Levy 1985, B\&C - Burkhart \& Coupry 1989 } 

\enddata
\end{deluxetable}